\documentclass[aps,prl,twocolumn,showpacs,showkeys]{revtex4-1}%endfloats
\usepackage{graphicx}%
\usepackage{dcolumn}
\usepackage{amsmath}   

\usepackage{bm}
\usepackage{longtable}

\begin{document}

\title{Cavity-water interface is polar}
\author{Allan D.\ Friesen} 
\author{Dmitry V.\ Matyushov}
\affiliation{Center for Biological Physics, Arizona State University, 
PO Box 871604, Tempe, AZ 85287-1604                   }
%\email{dmitrym@asu.edu}
\begin{abstract}
  We present the results of numerical simulations of the
  electrostatics and dynamics of water hydration shells surrounding
  Kihara cavities given by a Lennard-Jones (LJ) layer at the surface
  of a hard-sphere cavity. The local dielectric response of the
  hydration layer substantially exceeds that of bulk water, with the
  magnitude of the dielectric constant peak in the shell increasing
  with the growing cavity size. The polar shell propagates into bulk
  water to approximately the cavity radius. The statistics of the
  electrostatic field produced by water inside the cavity follow
  linear response and approach the prediction of continuum
  electrostatics with increasing cavity size.
\end{abstract}

\maketitle

Nanoscale interfaces of polar liquids combine strong distortions of
the liquid density profile with highly perturbed long-range
electrostatic correlations. The interfacial density, and the related
diffusional dynamics, are governed by short-range packing restrictions
and are relatively short-ranged. Nevertheless, the density
fluctuations of the interfacial region are critical for the long-range
hydrophobic forces \cite{Ball:08,ChandlerNature:05} and the related
weak dewetting of interfaces of non-polar solutes
\cite{HummerPRL:98,Berne:09}. In contrast, electrostatic interactions,
and the orientational correlations of multipolar moments, are
long-ranged. They are assigned macroscopic length-scale in the Maxwell
(continuum) electrostatics propagating the effect of partial charges
at dielectric interfaces on the length-scale of the Coulomb
potential. Whether this picture is correct for polar liquids and how
the surface polarization is screened by the mobile liquid dipoles
remains an open question \cite{DMepl:08}, the resolution of which will
define the limits of continuum electrostatics in application to
nanoscale interfaces of molecular liquids.

Water presents a particular challenge to the problems of interfacial
dynamics and thermodynamics since energetically strong hydrogen bonds
add a short-range scale competing with long-range electrostatic
forces.  The properties of hydration layers surrounding nanoscale
solutes indeed turn out to be unusual. Apart from ubiquitous
hydrophobic interactions linked to the structure of the water
interface \cite{Ball:08}, measurements of microscopic electrostatics
of the protein/water interface have shown some surprising results.
Electrostatics on the microscopic scale is traditionally probed by the
dynamic and static band-shifts of optical dyes \cite{Jimenez:94}.  The
corresponding Stokes-shift dynamics at the protein/water interface
showed a long exponential decay absent for the free chromophores in
solution. This observation has prompted the label of ``biological
water'' for hydration layers of biopolymers \cite{Pal:04}. While the
cause of this effect is still debated \cite{Zhang:07,Nilsson:05},
various extent of slowing of the collective Stokes shift dynamics has
been universally observed at protein/water interfaces
\cite{Pal:04,Zhang:07}. Further, the hydration shells around proteins
were found to carry high local polarity \cite{DMpre2:08}, thus linking
the slower dynamics to the structural reorganization of water in the
form of a polarized cluster around proteins
\cite{Ebbinghaus:07}. Unfortunately, the problem of the protein
hydration is inseparable from the complex protein dynamics
\cite{Frauenfelder:09}. Studies excluding this latter component are
therefore necessary to understand the electrostatics of hydration
layers when the solute size grows to the nanoscale. This is the goal
of this report.

Here we present extensive Molecular Dynamics (MD) simulations of the
structure of water around non-polar solutes (cavities). In contrast to
previous active research in this field
\cite{ChandlerNature:05,HummerPRL:98,Berne:09,GardePRL:09}, we ask here the
following questions: (i) how polar is the interface? (ii) how far into
the bulk does the polarity perturbation propagate? and (iii) how are
the orientational dipolar dynamics of the hydration layers affected by
the solute? The main result of this study is the observation of a
significant increase of the local water polarity at the interface,
with the region of enhanced polarity extending into the bulk to
approximately the cavity radius.

The water nanoscale interface was modeled by inserting spherical
solutes carrying a hard-sphere (HS) core surrounded by a Lennard-Jones
(LJ) potential layer. The interactions with the SPC/E oxygen is
then given by the Kihara solute-solvent potential
\begin{equation}
  \label{eq:2}
  \phi(r) = 4\epsilon_{\text{LJ}}\left[\left(\frac{\sigma}{r-r_{HS}}\right)^{12} 
            - \left(\frac{\sigma}{r-r_{HS}}\right)^6\right].
\end{equation}
The LJ well has the width $\sigma=3$ \AA\ and the energy $\epsilon_{\text{LJ}}$ for
which two values were used: $\epsilon_{\text{LJ}} = 0.65\mbox{ kJ/mol}$ equal
to the LJ energy between oxygens of SPC/E water and $\epsilon_{\text{LJ}} =
20\mbox{ kJ/mol}$ close to the energy of hydrogen bonds in bulk water.
The cavity size was varied by changing the HS radius $r_{\text{HS}}$
in the range 0--12 \AA.  The number of waters in the simulation cell was
varied to allow sufficiently large solvation layers, with 4053 and
11845 hydration waters used for the smallest and largest solutes,
respectively.  The trajectories for analysis were 5 ns long, following
100--500 ps equilibration.  Simulations were performed with cubic
periodic boundary conditions, at 273 K and zero pressure, with
Berendsen thermostat and barostat and a timestep of 2 fs.  Ewald sums
with tin foil boundary conditions were used for electrostatic
interactions.

A spherical solute induces a spherical symmetry breaking in an
otherwise isotropic liquid. The orientational structure of the
interface consistent with this imposed symmetry is characterized by
the first- and second-order orientational order parameters: $p_1(r) =
(N(r))^{-1}\left\langle \sum_{r_j<r } \mathbf{\hat r}_j \cdot \mathbf{\hat
    m}_j\right\rangle $
and $p_2(r)=(2N(r))^{-1}\left\langle \sum_{r_j<r} [3 (\mathbf{\hat r}_j \cdot
\mathbf{\hat m}_j)^2 -1]\right\rangle $. These parameters project the unit dipolar
vectors $\mathbf{\hat m}_j$ within the shell of radius $r$ on the
radial direction $\mathbf{\hat r}_j=\mathbf{r}_j/ r_j$, $N(r)$ is the
number of waters within the shell.  The first hydration layer is then
defined as $R\leq r \leq  R + 1.5$ \AA, $R=r_{\text{HS}}+\sigma$ and the
corresponding order parameters are $p_1^I$ and $p_2^I$.

\begin{figure}
  \centering
  \includegraphics*[width=6cm]{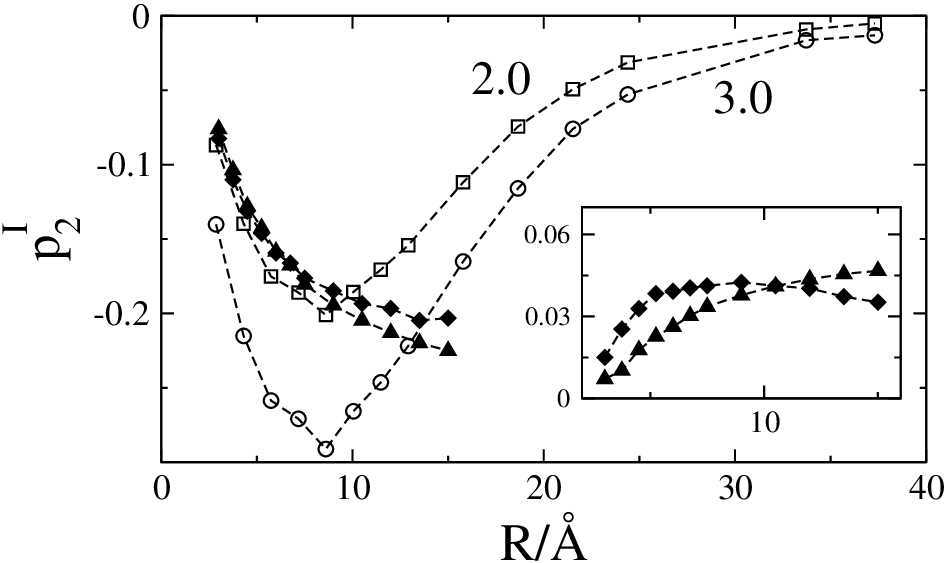} 
  \caption{The second-order orientational order parameter $p_2^I$ of
    the water first shell vs the cavity radius $R=r_{\text{HS}} + \sigma $.
    The solid diamonds and triangles refer to cavities in water with
    $\epsilon_{\text{LJ}}=0.65$ and 20 kJ/mol, respectively. The open points refer to
    cavities in the fluid of dipolar hard spheres \cite{DMepl:08} with
    the reduced dipole moments $(m^*)^2=\beta m^2/ \sigma_s^3$ equal to 2.0
    (open squares) and 3.0 (open circles); $m$ is the dipole moment
    and $\sigma_s$ is the hard-sphere diameter of the solvent.  The
    inset shows the orientational parameter $p_1^I$ of the first
    hydration layer. }
  \label{fig:1}
\end{figure}

The dielectric constant of a cavity-water mixture can be defined in
terms of the volume occupied by the cavity relative to the volume of
water \cite{Scaife:98}. The solute-solvent response function
describing the interfacial polarization can then be
calculated by accounting for dipolar fluctuations accumulated within a
radial water layer surrounding the cavity $\chi(r) = \beta\langle(\delta
\mathbf{M}(r))^2\rangle/(3V(r))$; $\beta=1/(k_{\text{B}}T)$ is the inverse
temperature. The water dipole moment $\mathbf{M}(r)$ in this equation
is taken over the radial shell between the spherical cavity 
and radius $r$ extending from the cavity center to the
bulk, $V(r)$ is the shell volume. The limit of infinite dilution
yields the bulk dielectric susceptibility of water $\chi =
\chi(\infty)=(\epsilon-1)/(4\pi)$, where $\epsilon$ is the water dielectric constant.

The fluctuation susceptibility $\chi(r)$ determines the dielectric
constant $\epsilon(r)$ accumulated within the radial layer. For simulations
employing periodic boundary conditions with tin-foil boundary around
replicas of the simulation cell the connection between the simulated
variance of the dipole moment and the dielectric constant is
particularly simple \cite{Neumann:86}: $\epsilon(r) = 1 + 4\pi\chi(r)$. The
macroscopic dielectric constant of water is then $\epsilon = \epsilon(\infty)$.

The orientational structure of polar liquids at interfaces is strongly
affected by short-range orientational correlations.  Unsaturated
hydrogen bonds of surface waters produce preferential in-plane
orientations of water's dipoles \cite{Lee:84} as reflected by the
first and second orientational order parameters (Fig.\
\ref{fig:1}). The first-order parameter $p_1^I$ is nearly zero
pointing to no preferential radial orientation (inset in Fig.\
\ref{fig:1}), while $p_2^I$ is non-zero and negative, in accord with
the preferential in-plane orientation of the dipoles. This
orientational pattern is specific for water and does not necessarily
repeat itself in other polar liquid.  For comparison, $p_2^I$ of
hard-sphere dipoles at the surface of a spherical cavity
\cite{DMepl:08} passes through a minimum (Fig.\
\ref{fig:1}). Orientational order first grows with increasing cavity
size as frustrations of dipolar orientations are released with the
growing number of dipoles. However, further increase of the cavity
size leads to a weak dewetting of the interface by the puling force of
the liquid \cite{HummerPRL:98} with the resulting destruction of the
interfacial orientational order. In contrast, water preserves its
parallel interfacial order, mostly determined by its hydrogen-bond
network and not much effected by the strength of the solute-solvent LJ
attraction (Fig.\ \ref{fig:1}).

\begin{figure}
  \centering
  \includegraphics*[width=6cm]{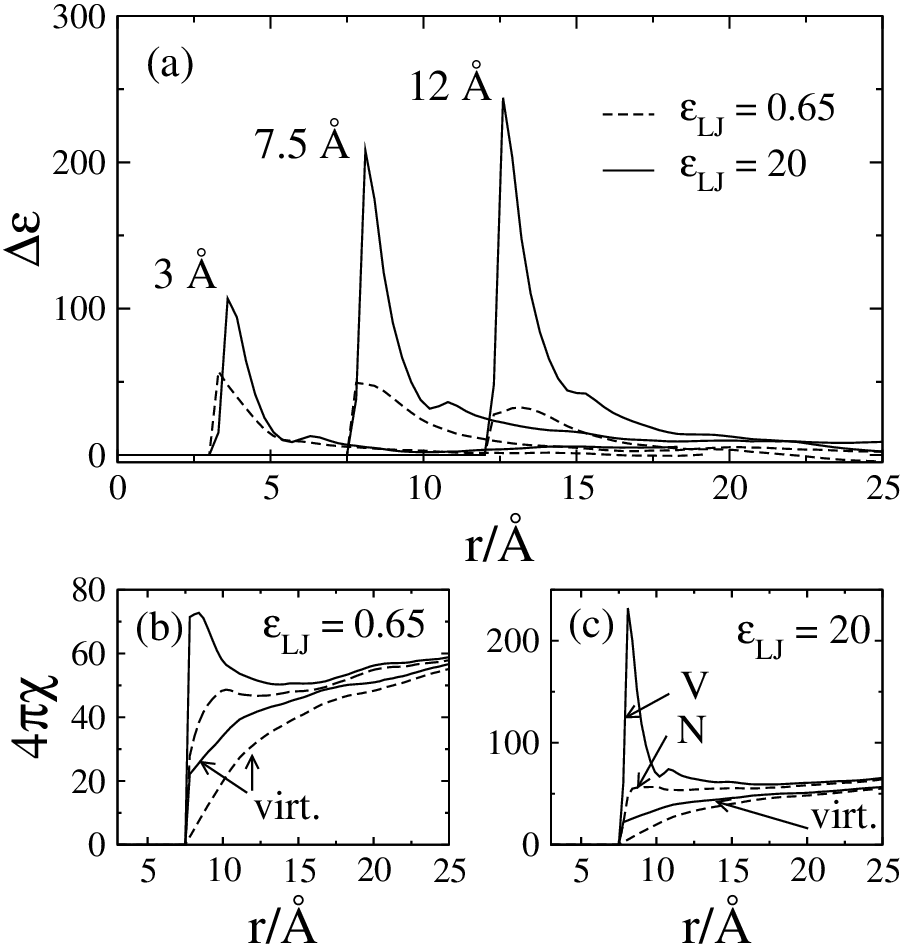}
  \caption{Dielectric constant of the hydration layer relative to the
    dielectric constant of the same layer around a virtual cavity for
    three cavity sizes indicated in the plot (a).  The solid and
    dashed lines refer to $\epsilon_{\text{LJ}}=20$ and 0.65 kJ/mol,
    respectively. The response functions defined through the volume of
    the shell (``V'', solid lines) and through the number of shell
    waters (``N'', dashed lines) are shown in (b) for
    $\epsilon_{\text{LJ}}=0.65$ kJ/mol and in (c) for $\epsilon_{\text{LJ}}=20$
    kJ/mol; $R = 7.5$ \AA.  $\chi(r)$ for the virtual cavity is marked as
    ``virt.'' }
  \label{fig:2}
\end{figure}

The function $\epsilon(r)$ calculated for shells around cavities is compared
to the same function calculated from shells around Lorentz's virtual
cavity \cite{Scaife:98} (water molecules between radii $R$ and $r$)
taken from configurations of pure water without cavity inserted. The
difference of the two functions shows a sharp peak (Fig.\
\ref{fig:2}a) pointing to an effectively higher polarity of hydration
shells around cavities compared to shells in bulk water.

In order to distinguish between the orientational and density origins
of the peak in the dielectric constant, we have plotted in
Figs.\ \ref{fig:2}(b,c) $\chi(r)$ defined as above in comparison with the
susceptibility normalized to the number of waters in the shell
$N(r)$: $\chi_N(r)=\beta \rho \langle (\delta \mathbf{M}(r))^2\rangle/(3 N(r))$, where $\rho$ is the
number density of bulk water. This comparison shows that the origin of
$\Delta \epsilon$ is a composite effect of changes in both the local density and
orientational structure. Even though a significant part of $\Delta \epsilon$ comes
from the increased density in the first solvation layer, particularly
at the large solute-solvent LJ attraction (Fig.\ \ref{fig:2}(c)), the
effect cannot be cast in terms of $N(r)$ only.

\begin{figure}
  \centering
  \includegraphics*[width=6cm]{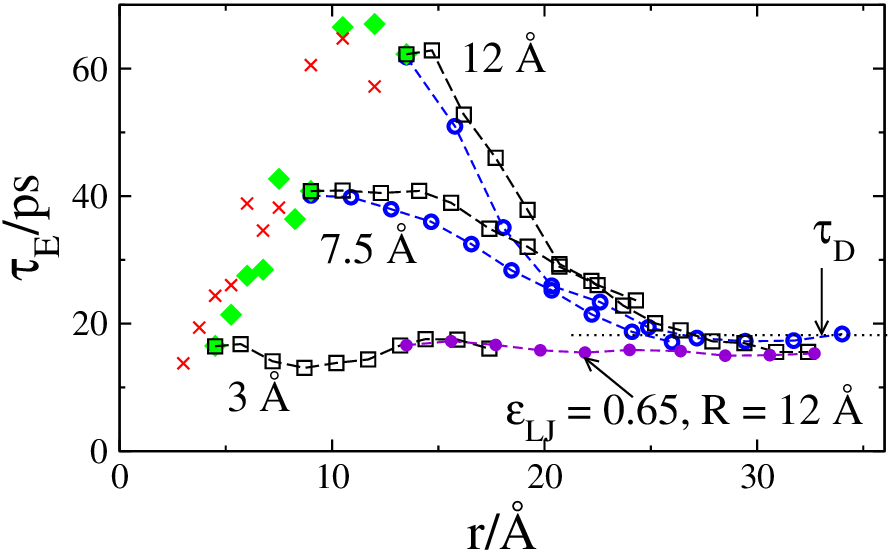}
  \caption{Exponential relaxation time of $\chi^I(t)$ (closed diamonds),
    $C_E(r,t)$ (open circles), and $\chi(r,t)$ (open squares);
    $\epsilon_{\text{LJ}}=20$ kJ/mol. Also shown is the exponential
    relaxation time of the self-correlation function of the unit
    vector $\mathbf{\hat e}^I(t)$ representing the dipole moment of
    the first-shell waters.  Different cavity sizes for the
    $r$-dependent relaxation times are indicated in the plot.  The
    filled circles refer to $R=12$ \AA\ and $\epsilon_{\text{LJ}}=0.65$ kJ/mol
    and the horizontal dotted line indicates the Debye relaxation time
    $\tau_D$ of pure SPC/E water.  The dashed lines in the plot are
    connecting the points. }
  \label{fig:3}
\end{figure}

Given the long range of the interfacial orientational order one
wonders how the dynamics of hydration layers are affected. We have
looked at several correlation functions. $\chi^I(t)=\beta \langle
(\delta\mathbf{M}^I(t)\cdot\delta \mathbf{M}^I(0)\rangle /(3V^I)$ is the time
self-correlation function of the dipole moment of the first solvation
layer and $\chi(r,t)$ is a similar correlation function extended to a
layer within the radius $r$ from the cavity's center. In addition, we
have calculated the correlation function $C_E(t)=\langle \delta
\mathbf{E}_s(r,t)\cdot\delta \mathbf{E}_s(r,0)\rangle $ of the electric field
$\mathbf{E}_s(r,t)$ produced at the cavity's center by the waters
within the $r$-shell. This latter correlation function represents the
Stokes shift dynamics of dipolar chromophores
\cite{Pal:04,Zhang:07}. All correlation functions were fitted to a sum
of a ballistic Gaussian decay and an exponential tail
\cite{Jimenez:94}. Figure \ref{fig:3} presents the compilation of the
results for the exponential relaxation time $\tau_E$.

The main observation from the dynamics calculations is a significant
growth of the exponential relaxation time of the first solvation layer
with the cavity size (filled diamonds in Fig.\
\ref{fig:3}). Consistent with the results for $\Delta \epsilon(r)$, this dynamics
perturbation propagates into the bulk to at least the distance of the
cavity radius.  The exponential decay time of $C_E(r,t)$ retraces the
corresponding time from $\chi(r,t)$.  The dipolar dynamics of the first
solvation layer are dominated by rotations of the dipole moment
$\mathbf{M}^I$, instead of its magnitude fluctuations, as is seen from
the self-correlation function of the unit vector $\mathbf{\hat
  e}^I(t)=\mathbf{M}^I(t)/ M^I(t)$ (crosses in Fig.\
\ref{fig:3}). This observation points to a high level of orientational
cooperativity in the first solvation layer, which does not decorrelate
by individual dipole rotations and instead rotates slowly as a
correlated dipolar domain. This dynamical slowing is however seen only
for the larger LJ attraction, $\epsilon_{\text{LJ}}=20$ kJ/mol, and no effect
of the solute on the dynamics is observed when the solute-solvent LJ
potential is similar to that in water, $\epsilon_{\text{LJ}}=0.65$ kJ/mol
(closed points in Fig.\ \ref{fig:3}). This result might help to
explain the conflicting literature \cite{He:07} on the subject of the
surface-induced alteration of the liquid dynamics.  The outcome seems
to be controlled by the strength of the solute-solvent interactions
and thus the surface composition.
 
Our results partially support Onsager's concept of ``inverted
snowball'' dynamics \cite{Onsager:77}. It stipulates that solvation
dynamics are slower close to a newly created charge compared to more
distant layers, ranging between the one-particle (slow) orientational
diffusion and the dielectric (fast) relaxation of the bulk.  The data
in Fig.\ \ref{fig:3} indeed show a speedup of dipolar relaxation into
the bulk.  However, this effect strongly depends on both the cavity
size and the solute-solvent LJ interaction.  It is expected to be
essentially absent for typical optical probes \cite{Jimenez:94}
consistent in size with the smallest cavity studied here. Further,
even for larger cavities, the slow dynamics of the closest solvation
layers are almost lost when the hydration layer is grown to the
boundaries of the simulation cell. At that point, the relaxation time
becomes the Debye relaxation time of bulk water (Fig.\
\ref{fig:3}). This observation implies that dipolar optical probes
placed inside the cavity \cite{Pal:04,Zhang:07,Nilsson:05} will not
pick up the slowing of the closest hydration shells and instead will
average the effect out by the electric field contributions from more
distant layers.
  
The statistics of dipolar interfacial fluctuation can be probed by the
chemical potential of electrostatic solvation, i.e.\ the free energy
of interaction of the charges inside the cavity with the surrounding
water solvent. We found that the dipolar field $\mathbf{M}(r)$ is
Gaussian and thus the linear response approximation should be
applicable.  The solvation chemical potential of a charge $\mu_q$ or a
dipole $\mu_d$ can then be found from the variance of the electrostatic
potential $\phi_s$ or the electric field $\mathbf{E}_s$ produced by the
solvent at the position of the corresponding multipole
\cite{Ben-Amotz:05}: $\mu_q = - \beta (q_0)^2 \langle (\delta\phi_s)^2\rangle/2$, $\mu_d = - \beta
(m_0)^2 \langle(\delta \mathbf{E}_s)^2\rangle/6 $. Here, $q_0$ and $m_0$ are,
correspondingly, the charge and point dipole within the cavity. The
averages in these relations do not depend, in linear response, on
whether the corresponding multipoles are actually present inside the
cavity \cite{Ben-Amotz:05}. They can therefore be calculated from our
simulations with empty cavities providing insights into how these
solvation free energies scale with the size of the solute and whether
the limit of continuum electrostatics is reached.

\begin{figure}
  \centering
  \includegraphics*[width=6cm]{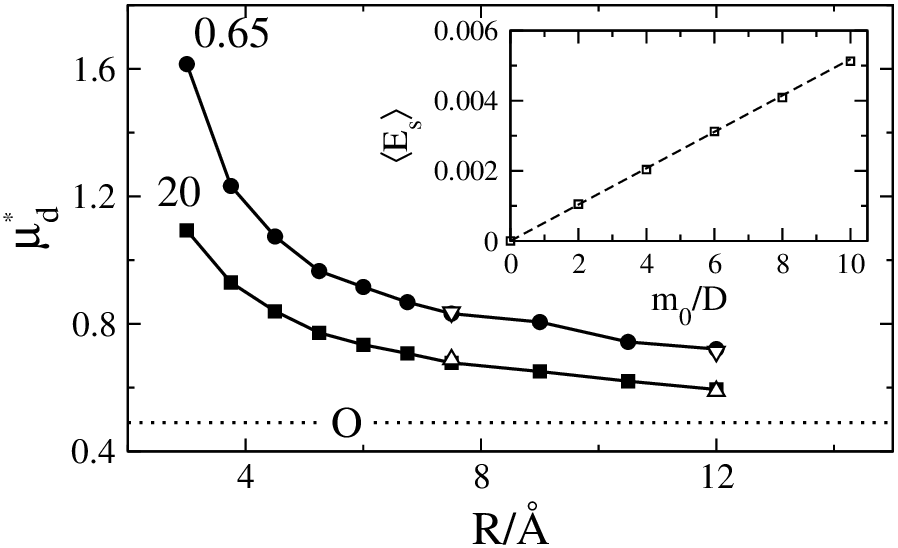}
  \caption{Chemical potential $\mu_d^*=\mu_d m_0^2/R^3$ of solvating the
    point dipole $m_0$ at the center of the cavity of radius $R$.  The
    solid points ($\epsilon_{\text{LJ}}=20$ kJ/mol, circles and 0.65 kJ/mol,
    squares) and obtained from the variance of the water electric
    field inside the empty cavity, $\mu_d \propto \langle (\delta E_s)^2\rangle$.  The open
    triangles are from the thermodynamic integration of the average
    electric field $\langle E_s\rangle$ produced by dipoles of increasing
    magnitude ($0<m_0<10$ D) positioned at the cavity center.  The
    average field (eV/D) is a linear function of the dipole magnitude
    $m_0$ (inset, $R=12$ \AA, $\epsilon_{\text{LJ}}=0.65$ kJ/mol).  The
    dashed line in the inset is the prediction based on the field
    variance inside the empty cavity (not the best fit). The dotted
    horizontal line is the result of continuum electrostatics given by
    the Onsager relation (``O'').  }
  \label{fig:4}
\end{figure}

We found noticeable effects of the size of the simulation cell on $\langle
(\delta\phi_s)^2\rangle$ and much smaller size effects on $\langle(\delta \mathbf{E}_s)^2\rangle$. We
have therefore chosen to look at the statistics of the field
fluctuations and the corresponding chemical potential of dipole
solvation. These results are summarized in Fig.\ \ref{fig:4} showing
$\mu_d^*=\mu_d R^3/(m_0)^2$. This dimensionless parameter is expected to
approach, with growing cavity size, the size-independent limit of
continuum electrostatics, given by the Onsager equation
\cite{Scaife:98} $\mu_d = (\epsilon -1)/(2\epsilon +1) \simeq 0.5$. The values of
$\mu_d$, although noticeably higher at intermediate sizes, indeed seem
to approach this limit. The open points in Fig.\ \ref{fig:4} are
obtained by thermodynamic integration of average energies of point
dipoles placed at the cavity center. A good agreement between $\mu_d$
from the field variance and from the thermodynamic integration, as
well as the linear dependence of $\langle E_s\rangle $ vs $m_0$ (inset in Fig.\
\ref{fig:4}), testifies to the validity of the linear response
approximation.

The chemical potential $\mu_d$ is not strongly affected by the
strength of the solute-solvent LJ attraction. The range of
$\epsilon_{\text{LJ}}$ values studied here probably covers most of
situations of practical interest. Deviations of the electric field
variance from the area between the two curves shown in Fig.\
\ref{fig:4} might therefore be used to identify nonlinear solvation
typically associated with hydrogen bonding between water
and the solute \cite{DMjpca1:06}.
% Such cases do exist as is shown by a point for a small charge-transfer
% solute (p-nitroanyline) which shows non-linear solvation
% \cite{DMjpca1:06} and indeed falls outside the linear response region
% between the two curves in Fig.\ \ref{fig:4}.

The results obtained here must have significant implications for
self-assembly of nano-sized objects and biological activity of
hydrated biopolymers. Polar solvation layers around solutes are
expected to screen the inside charges. In the crowded environment of a
living cell \cite{Douglas:09} this screening will reduce interactions
between multipolar solutes. The high-polarity layer is also
characterized by slower dipolar solvation. However, this effect is is
not picked up by the dynamics of dipolar probes placed inside the
cavity. A slow relaxation component observed in optical time-resolved
spectra \cite{Pal:04} therefore needs to be assigned to protein
motions pushing the hydration layers \cite{Nilsson:05}.

\acknowledgments This research was supported by the DOE, Chemical
Sciences Division, Office of Basic Energy Sciences (DEFG0207ER15908).

%\bibliography{chem_abbr,dielectric,dm,statmech,protein,liquids,solvation,dynamics,bioet,surface,photosynthNew}

%\end{document}

%Merlin.mbs v4.21 2009-07-09.
%

\end{document}